\documentclass[cits]{PoS}

\usepackage{natbib}
\setcitestyle{numbers,square} 
\usepackage{subfigure}

\def\deg{\hbox{$^\circ$}}
\newcommand{\mr}{\mathrm}
\newcommand{\nh}{\hbox{$N_{\rm H}$}}
\newcommand{\hcm}[1]{$\times 10^{#1}$ cm$^{-2}$}

\title{X-ray spectral and timing investigations of XTE J1752-223}

\ShortTitle{X-ray spectral and timing investigations of XTE J1752-223}

\author{{Holger Stiele}\\
        INAF Osservatorio Astronomico di Brera\\
        E-mail: \email{holger.stiele@brera.inaf.it}}

\author{Teo Mu\~noz-Darias \\
        INAF Osservatorio Astronomico di Brera\\
        E-mail: \email{tmd@brera.inaf.it}}

\author{Sara Motta\\
        INAF Osservatorio Astronomico di Brera\\
        E-mail: \email{sara.motta@brera.inaf.it}}

\author{Tomaso Belloni\\
       INAF Osservatorio Astronomico di Brera\\
        E-mail: \email{tomaso.belloni@brera.inaf.it}}

\abstract{We report on X-ray monitoring observations of the transient black hole candidate (BHC) XTE J1752-223 with the Rossi X-ray Timing Explorer (RXTE). The source was discovered on 2009 October 23 and during its low/hard state, which lasted for at least 25 days, all timing and spectral properties were similar to those of Cyg X-1 during its canonical hard state. 
The combined PCA/HEXTE spectra were well fitted by an absorbed broken powerlaw with a high energy cutoff. When RXTE observations were resumed, after an observational gap due to solar constraint, the source was in the hard intermediate state. The evolution through the hardness intensity diagram and the timing properties observed in the power density spectrum suggest that the source crossed all the canonical BHCs states. We discuss the different states and present the results of our spectral and timing investigations. }

\FullConference{25th Texas Symposium on Relativistic Astrophysics - TEXAS 2010\\
		December 06-10, 2010\\
		Heidelberg, Germany}

\begin{document}

\section{Introduction}
Black hole X-ray transients (BHTs) stay most of the time in quiescence. They represent the majority of the black hole binary (BHB) population known so far. During outburst BHTs show a characteristic evolution of their spectral and temporal properties. This led to the definition of different states: at the begin of the outburst the source is in the so-called low/hard state (LHS), it then evolves to the high/soft state (HSS) and finally returns to the LHS. Although this general behaviour is widely agreed on, the exact definition of the states and especially of the transition between these states are still under debate. In this work we follow the classification of \citep{2005Ap&SS.300..107H}  (see however \citep{2006csxs.book..157M} for an alternative classification and \citep{2009MNRAS.400.1603M} for a comparison). 
 
XTE J1752-223 was discovered by the Rossi X-ray timing explorer (RXTE) on 2009 October 23 \citep{2009ATel.2258....1M} at a 2 to 10\,keV flux of 30\,mCrab. A daily monitoring by RXTE to follow up the outburst evolution was triggered by 
significant similarities with the typical properties of a BHT during the low hard state (LHS) as well as detections of an optical and a radio counterpart.
An overview paper, including spectral and time variability studies, based on RXTE Proportional Counter Array (PCA) data, was presented by  \citep{2010ApJ...723.1817S}. A two day long RXTE observation taken in the early phase of the outburst was analysed by \citep{2010MNRAS.404L..94M}. The results obtained from MAXI GSC and \textit{Swift} were presented by \citep{2010PASJ...62L..27N} and \citep{2011MNRAS.410..541C}, respectively.

\section{Observations and data analysis}
\label{Sec_Obs}
We investigated 206 RXTE observations taken between 2009 October 26 and 2010 July 3, which cover the whole outburst, to present a comprehensive spectral-timing study of XTE J1752-223. In order to do so, we included data obtained by the High Energy X-ray Timing Experiment (HEXTE; 20\,--\,200\,keV) on board RXTE. This means, that we also investigate the high energy range above $\sim$45 keV compared to \citep{2010ApJ...723.1817S}. 
 
For our timing analysis, we used PCA channels 0 -- 35 (2\,--\,15\,keV) only. The PCA Standard 2 mode (STD2), which covers the 2\,--\,60\,keV range with 129 channels, was used for spectral analysis. Energy spectra were extracted from PCA and HEXTE data using the standard RXTE software within \textsc{heasoft} V.~6.9. From the PCA only Proportional Counter Unit 2 data were used. From HEXTE we used Cluster B data for observations taking before 2009 December 14. For later observations the ``on source'' spectrum was obtained from Cluster A, while the background spectrum was estimated using Cluster B data.\footnote{This was necessary as Cluster B stopped ``rocking'' on 2009 December 14 and observes now in ``off source'' position only . Cluster A has been ``staring'' in ``on source'' position since 2006 October 20.} To account for residual uncertainties in the instrument calibration a systematic error of 0.6 and 1 per cent was added to the PCA and HEXTE data, respectively. Nevertheless there are still additional residuals in the HEXTE spectra obtained after 2009 December 14. We will address this point in more detail in Sect.\,\ref{Sec_Spec}.

\section{Timing investigations}
\label{Sec_Timing}
\begin{figure}
\begin{center}
\includegraphics[scale=0.45]{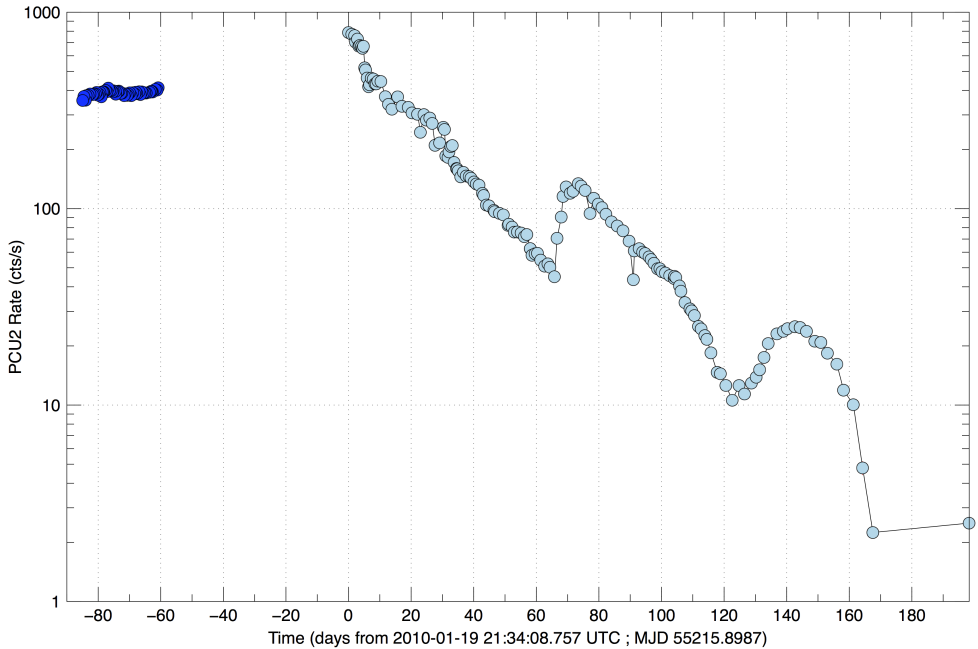}
\caption{The PCA light curve. The initial LHS of XTE J1752-223 is marked by the dark blue dots on the left hand side of the diagram. The following gap was due to solar constraint. The date of  T$=$0 is 2010-01-19 21:34:08.757 UTC; MJD 55215.8987.\@}
\label{lcurve}
\end{center}
\end{figure}

The Proportional Counter Array (PCA) light curve, using data of PCU \#2, is shown in Fig.\,\ref{lcurve}. The count rate is rather constant during the first part of the outburst (dark blue points in Fig.\,\ref{lcurve} at T$<-60$\,d), which corresponds to the initial LHS \citep{2010MNRAS.404L..94M}. During the following gap the source was not observable due to solar constraint. In the first observation taken after this gap the count rate has increased by  about a factor of two. From this point in time the source decreased in brightness, apart from tow periods of re-brightening.  

\begin{figure}
\begin{center}
\includegraphics[scale=0.45,angle=-90]{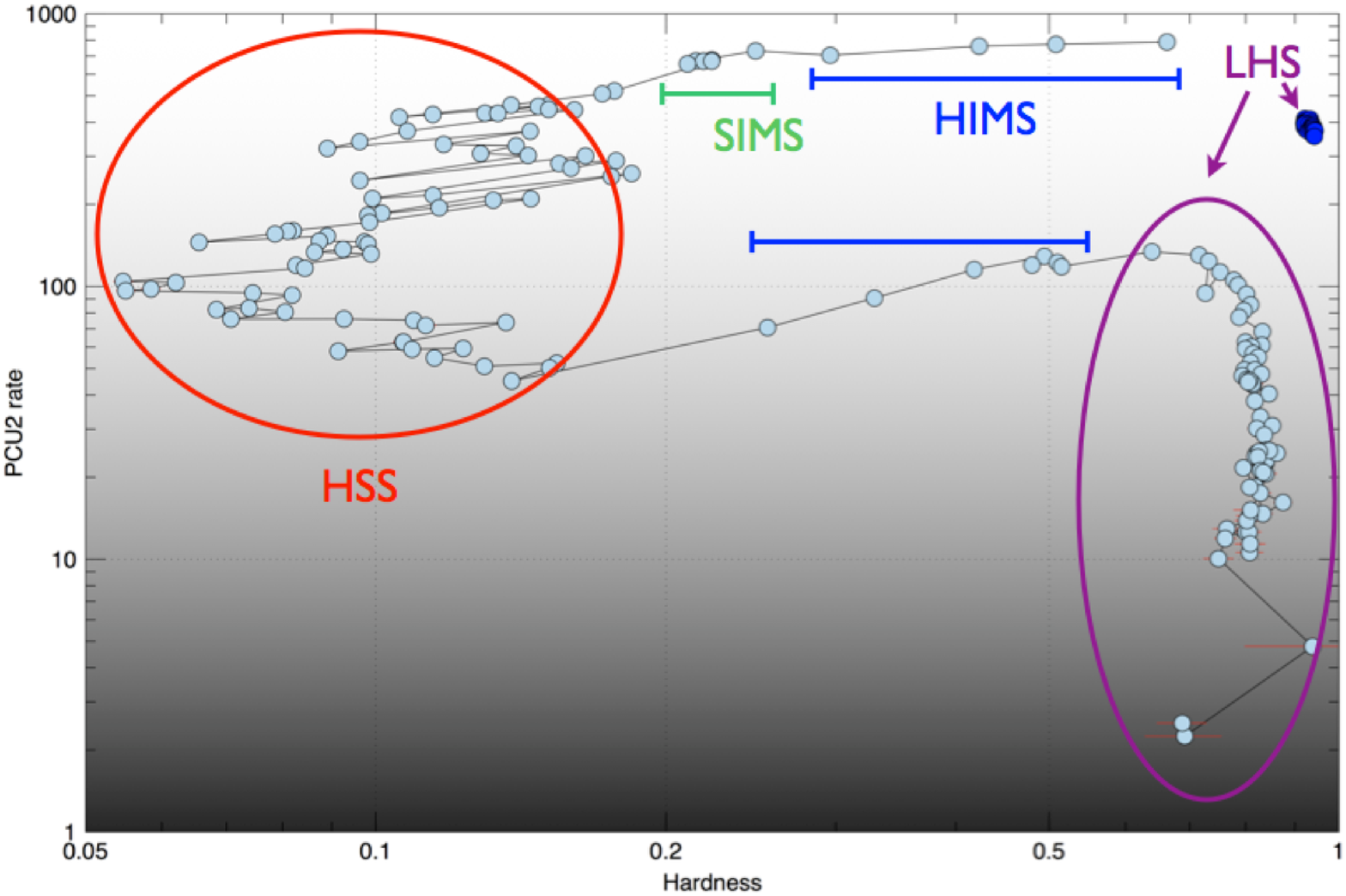}
\caption{The hardness intensity diagram. The different states (following the classification of [2]) through which XTE J1752-223 evolves are indicated  (see Sect.\,5).}
\label{HIdiag}
\end{center}
\end{figure}
\begin{figure}
\begin{center}
\includegraphics[scale=0.45,angle=-90]{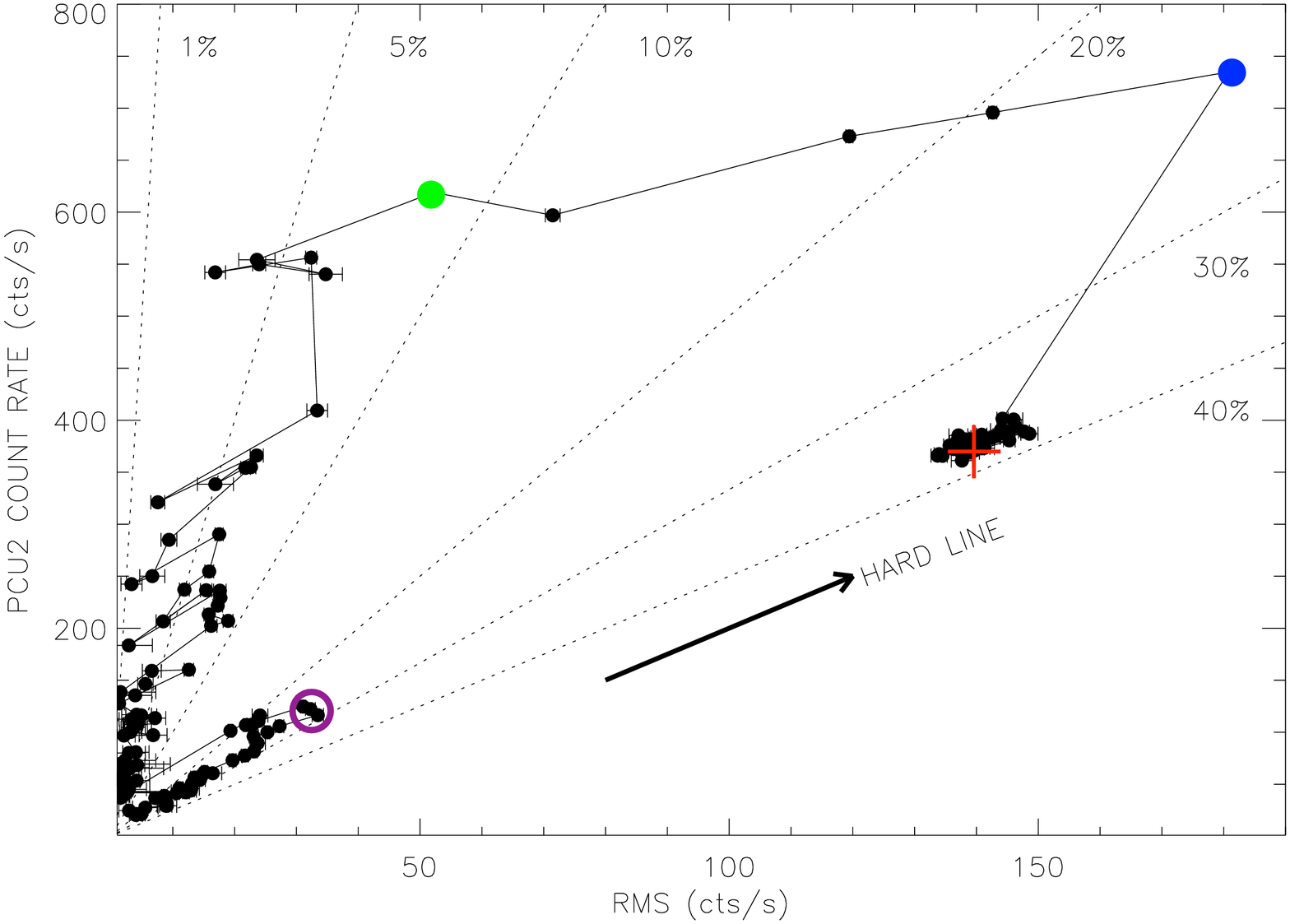}
\caption{The rms-intensity diagram gives the PCU2 count rate depending on the total rms. The LHS is located close to the line indicating a fractional rms of 40\%. The first observation is marked by a red cross. The onset of the HIMS is given by a blue dot, of the SIMS by a green dot. The onset of the LHS at lower luminosity is marked in violet. }
\label{rmsdiag}
\end{center}
\end{figure}

Figure \ref{HIdiag} shows the hardness intensity diagram (HID), which gives the PCU2 count rate depending on the hardness. XTE J1752-223 describes the standard q-shaped pattern; starting in the upper right corner (dark blue dots in Fig.\,\ref{HIdiag}) and evolving in counter clockwise direction.

The rms-intensity diagram (RID), which gives the PCU2 count rate depending on the total rms, is shown in Fig.\,\ref{rmsdiag}. 
It was introduced, based on GX 339-4 data, in a recent paper by \citep{2011MNRAS.410..679M}. This diagram allows to constrain different states without needing any spectral information.  All observations of the LHS are close to the line indicating a fractional rms of 40\%. The onset of the HIMS is given by a blue dot, of the SIMS by a green dot. The onset of the LHS at lower luminosity is marked in violet.  More information on the different states is given in Sect.\,\ref{Sec_States}.

\section{Spectral investigations}
\label{Sec_Spec}
We performed a broad-band spectral analysis by combining PCA (3\,--\,20\,keV) and HEXTE (20\,--\,200\,keV) data. The spectral fitting was done with \textsc{isis} V.~1.6.1 \citep{2000ASPC..216..591H}.

In a first step we tried to fit the PCA/HEXTE spectra with one-component models, such as a power law with cut-off or a multi-colour disc blackbody.  However, all models failed to describe the spectra properly (see also \citep{2010MNRAS.404L..94M}).  Following \citep{2010MNRAS.404L..94M}, the PCA/HEXTE spectra were fitted using an absorbed broken power law with a high energy cut-off. To account for the excess at 6.4\,keV a Gaussian centered at that energy was added. From day 2 onwards until day 68 an additional disc blackbody model was needed, representing the emission of the soft X-ray disc surrounding the black hole. The foreground absorption was fixed at \nh = 0.72\hcm{22} \citep{2010MNRAS.404L..94M}. 

As already mentioned in Sect.\,\ref{Sec_Obs}, all HEXTE observations taken after day 0 are affected by additional residuals, which are related to the fact that HEXTE detectors have stopped rocking. To take these residuals into account, we allowed the strength of the HEXTE background to be renormable (\texttt{corback} command in \textsc{isis}) during fitting and added three additional gaussians at the position of the strongest residuals (at $\sim$63\,keV, $\sim$53\,keV, and $\sim$40\,keV).
Nevertheless, some spectral fits still yielded unacceptable high values of $\chi^2_{\mr{red}}$ or totally un-physical parameter values. For these observations, we decided to model the HEXTE background during fitting, using a sophisticated model that takes known residual lines into account. 

\begin{figure}
\includegraphics[scale=0.9]{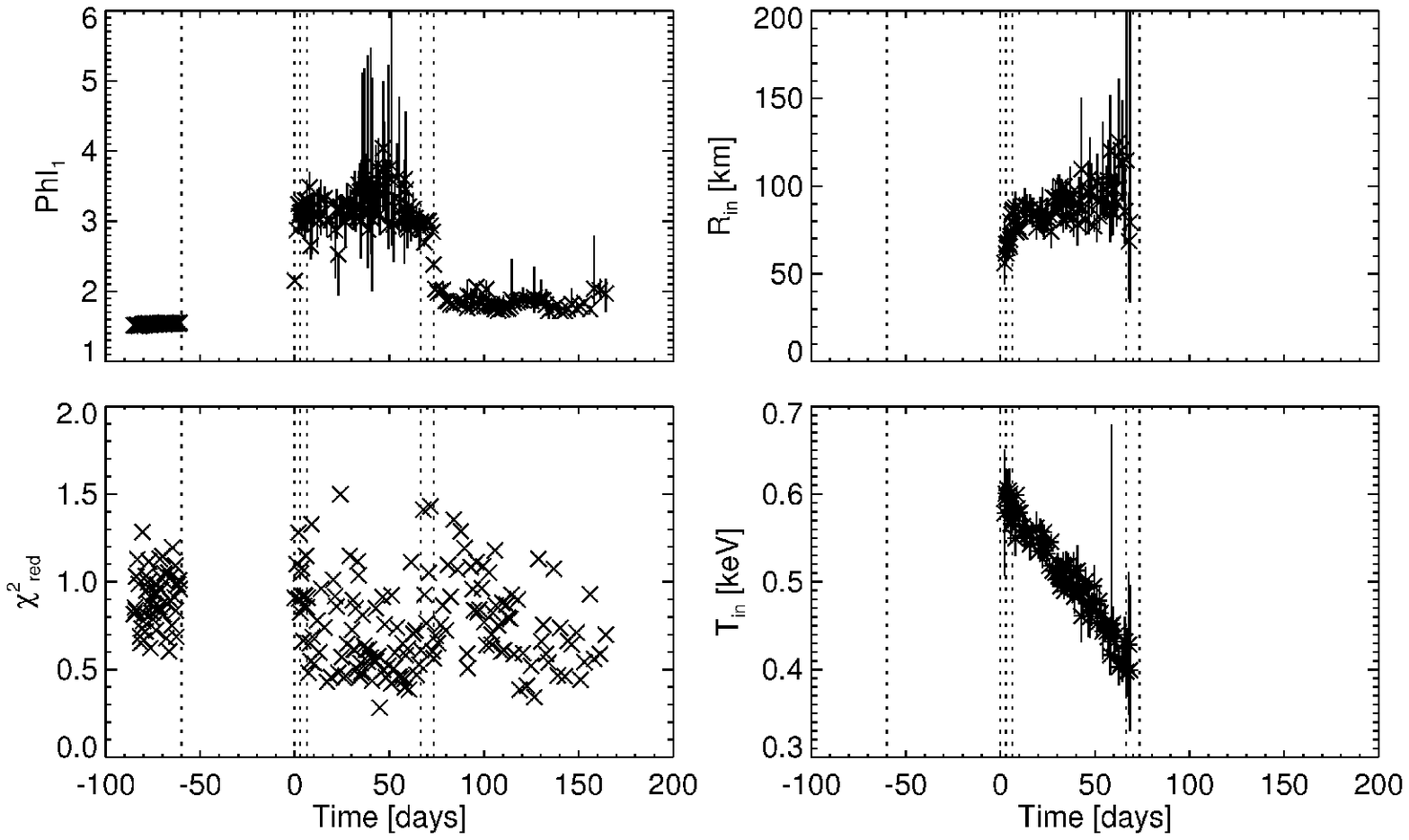}
\caption{Temporal evolution of selected spectral parameters. Given are the evolution of $\chi^2_{\mr{red}}$ (lower left panel), photon index below break energy (upper left panel), inner disk temperature (lower right panel), and inner disk radius (upper righty panel). The vertical dashed lines indicate times of (main) state transitions; from left to right: LHS, observation gap, HIMS, SIMS, HSS, HIMS, LHS (see also Sect.\,5). T$=0$ corresponds to 2010-01-19 21:34:08.757 UTC; MJD 55215.8987.}
\label{fig_specpar}
\end{figure}

The temporal evolution of $\chi^2_{\mr{red}}$ as well as of selected spectral parameters is given in Fig.\,\ref{fig_specpar}.
To derive the inner disc radius a distance of 3.5\,kpc  \citep{2010ApJ...723.1817S} and an inclination of 70\deg\  \citep{2010MNRAS.404L..94M} were assumed. The behaviour of spectral parameters during different states is presented in Sect.\,\ref{Sec_States}.

\section{The different states and their timing and spectral properties:}
\label{Sec_States}
\begin{itemize}
\item For the first 29 days (dark blue dots in Figs.\,\ref{lcurve}, \ref{HIdiag}; about 50 observations) the count rate was rather constant. During this time the source was in the low/hard state (LHS), showing rms variability of $\sim$40\% (see Fig.\,\ref{rmsdiag},  \citep{2010MNRAS.404L..94M}). The spectral components are rather constant during this state (see also Fig.\,\ref{fig_specpar}): fold energy (cut-off) $\approx$145\,keV, break energy $\approx$10\,keV, photon index below break (PhI$_1$) $\approx$1.53, photon index above break (PhI$_2$) $\approx$1.28.  
Furthermore they are  very similar to those of Cyg-X1 \citep{2010MNRAS.404L..94M}.
\item After that XTE J1752-223 was not observable with RXTE for a further 60 days due to solar constraint. 
\item When the source was observed again with RXTE its count rate has increased and the source was in the hard intermediate state (HIMS). During this observation and the following two observations the source showed type C QPOs (Quasi Periodic Oscillations) at 2.2\,Hz, 4.1\,Hz, and 5.5\,Hz, respectively, while the rms variability decreased from 25\% to 18\% (see Fig.\,\ref{rmsdiag}). The spectrum was softer than in the LHS, with PhI$_{1}\sim$2.8 and PhI$_{2}\sim$2.0. The high energy cut-off is no longer well constrained.
\item XTE J1752-223 evolved further through the soft intermediate state (SIMS), showing type A/B QPOs, and an rms variability of less than 10\%. 
In the following the source showed a main transition to the high/soft state (HSS) as well as several secondary transitions between the SIMS and HSS. A detailed discussion of these transitions will be given in a forthcoming paper. 
With the transition to the SIMS R$_{\mr{in}}$ was $\sim$60\,km and T$_{\mr{in}}$ was $\sim$0.6\,keV. During the HSS R$_{\mr{in}}$ increased slightly, while T$_{\mr{in}}$ decreased continuously. 
\item After a further 59 days XTE J1752-223 passed through another HIMS at lower luminosity. During this transition T$_{\mr{in}}$ as well as PhI$_1$ decreased rapidly. 
\item Finally the source entered into the LHS again at lower luminosity. The spectral components are rather similar to those at the beginning of the outburst, apart from the fold energy, which cannot be well constrained. This is partly due to the source being fainter, but even more due to the large uncertainties in the HEXTE spectra.
\item In total XTE J1752-223 was for more than about 300 days in outburst and evolved through all canonical BHCs states, before it faded into quiescence again.
\end{itemize}

\acknowledgments
The research leading to these results has received funding from the European Community's Seventh Framework Programme (FP7/2007-2013) under grant agreement number ITN 215212 "Black Hole Universe"

\bibliographystyle{plainnat}
\bibliography{/Users/apple/work/papers/my2010}



\end{document}